\documentstyle [twocolumn,epsf,amssymb]{mn}
\newcommand{\mincir}{\raise
-3.truept\hbox{\rlap{\hbox{$\sim$}}\raise4.truept\hbox{$<$}\ }}
\newcommand{\magcir}{\raise
-3.truept\hbox{\rlap{\hbox{$\sim$}}\raise4.truept\hbox{$>$}\ }}
\newcommand{\minmag}{\raise
-3.truept\hbox{\rlap{\hbox{$<$}}\raise5.truept\hbox{$<$}\ }}
\newcommand{\be}{\begin{equation}}
\newcommand{\ee}{\end{equation}}

\newcommand{\ba}{\begin{eqnarray}}
\newcommand{\ea}{\end{eqnarray}}
\newcommand{\brr}{\begin{array}}
\newcommand{\err}{\end{array}}
\newcommand{\bc}{\begin{center}}
\newcommand{\ec}{\end{center}}

\title[growth index and dark energy clustering]
{The growth index of matter perturbations using the clustering of dark energy}
\author[Spyros Basilakos]{Spyros Basilakos$^{\star}$\\
\vspace{0.1cm}$^{\star}$ Academy of Athens, Research Center for Astronomy \& Applied
  Mathematics, Soranou Efessiou 4, 11-527, Athens, Greece}

\begin{document}
  
\maketitle

\begin{abstract}
We have put forward a new unified framework
which provides a consistent and rather complete account of the
growth index of matter perturbations 
in the regime where the dark energy is allowed to have clustering.
In particular, we find that the growth index 
is not only affected by the cosmological parameters
but rather it depends on the choice of the considered 
dark energy (homogeneous or clustered).
Using the {\em Planck} priors and performing a standard $\chi^2$-minimization 
between theoretical expectations and growth data, we statistically 
quantify the ability of the growth index to represent the observations. 
Finally, based on the growth index analysis 
we find that the growth data favor the clustered dark energy scenario.

\end{abstract}

\vspace{1.0cm}

\section{Introduction}
The available high-quality cosmological observational data (e.g. supernovae type
Ia, CMB, galaxy clustering, etc), accumulated during the last two decades,
have enabled cosmologists to gain substantial confidence that modern
cosmology is capable of quantitatively reproducing the details of
many observed cosmic phenomena, including the late time
accelerating stage of the Universe. A variety of studies
have converged to a cosmic expansion history involving a spatially
flat geometry and a cosmic dark sector formed by cold dark matter and some
sort of dark energy (DE), endowed with large negative pressure, in order to
explain the observed accelerating expansion of 
the Universe (cf. Komatsu et al. 2011; Blake et al. 2011; Hinshaw et al. 2013; 
Farroq, Mania \& Ratra 2013; Spergel et al. 2013; 
Ade et al. 2014 and references therein).

Although there is a common agreement concerning the basic 
ingredients of the universe, there are different ideas 
regarding the underline physical mechanism which is responsible
for the cosmic acceleration. 
These patterns are based either on the existence of new fields in nature (DE)
or in some modification of Einstein's general relativity,
with the present accelerating stage appearing as a sort of geometric effect.
In order to test the latter possibilities,
it has been proposed that measuring the so called 
growth index, $\gamma$, could provide an efficient
tool to discriminate between modified gravity models and DE 
models which adhere to general relativity 
(Linder \& Cahn 2007; Nesseris \& Perivolaropoulos 2008; Wei 2008; 
Polarski \& Gannouji 2008; Gong, Ishak \& Wang 2009; Fu, Wu, \& Yu 2009; 
Tsujikawa et al. 2009; Basilakos \& Pouri 2012; 
Basilakos, Nesseris \& Perivolaropoulos 2013; Nesseris et al. 2013; 
Pouri, Basilakos \& Plionis 2014; Steigerwald, Bel, \& Marinoni 2014
and references therein). 

It is interesting to mention that 
most of the attempts towards estimating the growth 
index have a common basis, 
namely at the sub-horizon scales the DE component 
is expected to be smooth and thus we 
consider perturbations only on the matter component of the
cosmic fluid. 
But how would the growth index predictions change in the case where 
the DE is allowed to cluster?
In other words {\em can we estimate the growth index of matter fluctuations 
within the framework of clustered DE (hereafter CLDE)?} Indeed, this is an 
interesting question because an affirmative answer opens naturally a new path 
towards understanding the structure formation mechanism in the DE 
regime.

Recently, there are efforts towards investigating the linear 
growth of matter perturbations
in the context of CLDE (Bean \& Dore 2004; Ballesteros \& Riotto 2008; 
Sapone \& Kunz 2009; Sapone \& Majerotto  2012;
Batista \& Pace 2013; Dosset \& Ishak 2013; Batista 2014; 
Steigerwald, Bel \& Marinoni 2014; 
Pace, Batista \& Del Popolo 2014; Basse et al. 2014).
From the observational viewpoint, although it is difficult 
to measure the DE perturbations, it has been found that the 
homogeneous DE scenario fails to reproduce 
the observed concentration parameter of the massive galaxy 
clusters (Basilakos, Sanchez \& Perivolaropoulos 2009).
Similar results also found recently by 
Mehrabi, Malekjani \& Pace (2015) 
who claim that the CLDE models fit better the growth 
data than the homogeneous DE models (see also Nesseris \& Sappone 2015).
Generally, it has been shown that 
in order to search for observational effects of DE clustering 
we need to use the growth of matter fluctuations  
(Sapone, Kunz \& Amendola 2010).


The layout of the current paper is the following.
In section 2, we provide for the first time 
(to our knowledge) 
the growth index of matter fluctuations as a function of the DE perturbations. 
In section 3, we compare the theoretical predictions 
with the growth data in order to check the range of validity 
of the CLDE models and we discuss the variants
from the homogeneous case.
Finally, we summarize our results in section 3 

\section{Dark energy perturbations and the growth index}
Let us first 
present the basic linear equations that govern the evolution of the matter
and DE perturbations. Following the notations of  
Abramo et al. (2007; 2009) we write the following system
\be
\label{odedelta}
\ddot{\delta}_{m}+ 2H\dot{\delta}_{m}=
\frac{3H^{2}}{2}\left[\Omega_{m}\delta_{m}+\Omega_{de}\delta_{de}(1+3w)\right],
\ee
\be
\label{dodedelta}
\ddot{\delta}_{de}+(2H-\frac{\dot{w}}{1+w})
\dot{\delta}_{de}=
\frac{3(1+w)H^{2}}{2}\left[\Omega_{m}\delta_{m}+\Omega_{de}\delta_{de}(1+3w)\right]
\ee
where an overdot means cosmic time differentiation, 
$H(a)=H_{0}E(a)$ is the Hubble parameter, 
$w$ is the equation of state (EoS) parameter, 
$\Omega_{m}(a)=1-\Omega_{de}(a)=\Omega_{m0}a^{-3}/E^{2}(a)$,   
and $(\delta_{m},\delta_{de})$ denote the corresponding fluctuations.
We would like to stress that in order to obtain the above 
system Abramo et al. (2007) used the framework of the spherical collapse model
with the traditional top-hat approximation\footnote{Abramo et al. (2009) 
showed that Eqs. (\ref{odedelta}) and (\ref{dodedelta}) are valid also in the 
Post-Newtonian case (Lima, Zanchin \& Brandenberger 1997).}. Moreover, for 
simplicity these authors restrict their analysis to $c^{2}_{\rm eff}=w$ (where 
$c^{2}_{\rm eff}$ is the fluid's effective sound speed) assuming 
that the
EoS parameter remains the same either for the background expansion 
or for 
the spherical perturbations. In principle, since $c^{2}_{\rm eff}$ can take 
negative values one may expect that there are instabilities in the growth 
of DE perturbations. However, in the context of the spherical collapse model 
with the top-hat profile if we consider that 
$c^{2}_{\rm eff}$ is a scale independent quantity then it is easy to prove
that the pressure and the density gradients vanish which implies that 
DE instabilities do not exist (for more details see Abramo et al. 2007).
We will elaborate on the details of the general linear equations 
that lead the evolution of the matter
and DE perturbations in a forthcoming paper, but 
as a first step it is important to study the performance of the DE 
fluctuations in the most simple case.

For a constant EoS parameter [${\dot w}(a)=0$] 
the normalized Hubble function is written as 
\be
E(a)=\sqrt{\Omega_{m0}a^{-3}+\Omega_{de0}a^{-3(1+w)}}\;.
\ee
Let us first concentrate on Eq.({\ref{odedelta}) [for a more general analysis
see the appendix]. 
Inserting into the latter equation  
the growth rate of clustering $f=d{\rm ln}\delta_{m}/d{\rm ln} a$ and  
$\frac{d}{dt}=H\frac{d}{d\ln a}$, we derive after some algebra, that  
\be
\label{fzz222}
a\frac{df}{da}+f^{2}+\left( \frac{1}{2}-\frac{3}{2}w\Omega_{de}\right)f
= \frac{3}{2}\left[\Omega_{m}+ (1+3w)\Delta_{de} \Omega_{de} \right]
\ee
where $\Delta_{de}(a)\equiv \delta_{de}/\delta_{m}$ and 
$a(z)=(1+z)^{-1}$ is the scale factor. 
Notice that by definition 
in the case of the concordance $\Lambda$ cosmology the 
DE perturbations vanish implying $\Delta_{de}=0$.
Substituting the ansatz 
$f(\Omega_{m})\simeq \Omega^{\gamma(\Omega_{m})}_{m}$ (Peebles 1993; 
Wang \& Steinhardt 1998)
into Eq. (\ref{fzz222}), using a slow varying EoS parameter and 
performing simultaneously a first order Taylor expansion
around $\Omega_{m}(a)=1$ 
(for a similar analysis in the case of homogeneous DE 
see Refs. Linder \& Gahn; Nesseris \& Perivolaropoulos 2008; Gong et al. 2009; 
Tsujikawa, De-Felice \& Alcaniz 2013)
we find the following new approximate solution 
\be
\label{fzz333}
\gamma \simeq \gamma_{\rm HDE}+\frac{3\Delta_{de} (1+3w)}{6w-5}\Omega_{de} ,
\ee
where
\be
\gamma_{\rm HDE}  \simeq \frac{3(w-1)}{6w-5}+\frac{3(1-w)(2-3w)}{2(6w-5)^{2}(5-12w)}\Omega_{de} .
\ee
Obviously, from the above analysis it becomes evident that within the framework 
of the CLDE scenario  
the corresponding growth index is written 
in terms of the nominal growth index, namely $\gamma_{\rm HDE}$ 
(based on homogeneous DE)
plus an additional component which is related 
with the DE perturbations.
Of course, assuming a homogeneous dark energy ($\Delta_{de}=0$) the 
above 1st order solution reduces to the usual growth index functional form
(see Gong et al. 2009; Tsujikawa et al. 2009)  
as it should. Also, as it is expected at high enough 
redshifts ($z\gg 1$) since, $\Omega_{m} \simeq 1$ 
(or $\Omega_{de}\simeq 0$) we find that the 
asymptotic value of the growth index is not really affected 
by the clustering properties of the 
DE: $\gamma_{\infty}\approx 3(w-1)/(6w-5)$.
However, one may see that the CLDE could affect the growth index 
at intermediate redshifts. 
As an example, considering the {\it Planck} prior (Ade et al. 2014) 
($w=-1.13$) we find 
$\gamma\simeq 0.542+6.7\times 10^{-3}\Omega_{de}+
0.609 \Delta_{de} \Omega_{de}$, where $\Delta_{de}\sim {\cal O}(0.1)$ 
(see below).

Since, the solution (\ref{fzz333}) is valid at 
relative large redshifts one may want to treat 
the growth index also in the late universe
and indeed various candidates have been proposed in the literature.
In this paragraph we extend the original Polarski \& Gannouji (2008)
work for a general family of $\gamma(z)$ parametrization which is also valid
in the CLDE regime.
In particular, we phenomenologically parametrize $\gamma(z)$ 
by the following general relation
\be
\label{Param}
\gamma(z)=\gamma_{0}+\gamma_{1}y(z)\;.
\ee
Simply, the latter equation can be viewed as a first order Taylor expansion 
around some cosmological quantity such as $a(z)$ and $z$.
If we change variables in Eq.(\ref{fzz222})
from $a(z)$ to redshift [$\frac{df}{da}=-(1+z)^{-2}\frac{df}{dz}$]
and using $f(z)=\Omega_{m}(z)^{\gamma(z)}$ we obtain
\be
\label{Poll}
-(1+z)\gamma^{\prime}{\rm ln}(\Omega_{m})+\Omega_{m}^{\gamma}+
3w\Omega_{de}(\gamma-\frac{1}{2})+\frac{1}{2}=
\frac{3}{2}\Omega_{m}^{1-\gamma}X 
\ee
where prime denotes derivative with respect to redshift and 
\be \label{xxx}
X(z)=1+\frac{\Omega_{de}(z)}{\Omega_{m}(z)}\Delta_{de}(z)(1+3w)\;.
\ee 
Interestingly, for those $y(z)$ functions which obey $y(0)=0$ 
[or $\gamma(0)=\gamma_{0}$] one can write the 
parameter $\gamma_{1}$ in terms of $\gamma_{0}$. Indeed, 
at the present epoch 
[$z=0$, $\gamma^{\prime}(0)=\gamma_{1}y^{\prime}(0)$]
Eq.(\ref{Poll}) takes the form:
\be
\label{Poll2}
\gamma_{1}=\frac{\Omega_{m0}^{\gamma_{0}}+3w_{0}(\gamma_{0}-\frac{1}{2})
\Omega_{de0}+\frac{1}{2}-\frac{3}{2}\Omega_{m0}^{1-\gamma_{0}} X_{0}}
{y^{\prime}(0)\ln  \Omega_{m0}}\;.
\ee
Clearly, in the case of homogeneous DE ($\Delta_{de}=0$, $X=1$)
the above formula reduces to that of 
Polarski \& Gannouji (2008) for $y(z)=z$. Also, in the case of 
$y(z)=1-a(z)=\frac{z}{1+z}$ we fully recover literature results  
(Ishak \& Dosset 2009; Belloso, Garcia-Bellido \& Sapone 2011; 
Di Porto, Amendola \& Branchini 2012). 
Since, the formula $y(z)=z$ goes to infinity at large redshifts 
for the rest of the paper we concentrate on 
$y(z)=z/(1+z)$ (Ballesteros \& Riotto 2008) and thus $y^{\prime}(0)=1$.
Obviously, at large redshifts $z\gg 1$ we get 
$\gamma_{\infty}\simeq \gamma_{0}+\gamma_{1}$.    
Therefore, plugging $\gamma_{0}=\gamma_{\infty}-\gamma_{1}$ into
Eq.(\ref{Poll2}) and utilizing simultaneously 
$\gamma_{\infty}\approx 3(w-1)/(6w-5)$ we can derive the constants 
$\gamma_{0,1}$ in terms of $(\Omega_{m0},w,\Delta_{de0})$.

To this end, we would like to stress that the parametrization 
$f(a)\simeq \Omega_{m}(a)^{\gamma(a)}$ plays a significant role 
in structure formation studies since it greatly
simplifies the numerical calculations
of Eq.(\ref{odedelta}). Indeed, providing a direct integration 
of the above parametrization we easily find the linear growth factor
\begin{equation}
\label{Dz221}
\delta_{m}(a,\gamma)=a(z) \;{\rm exp} \left[\int_{a_{i}}^{a(z)} \frac{du}{u}\;
\left(\Omega_{m}^{\gamma}(u)-1\right) \right]
\end{equation}
where $a_{i}$ is the scale factor of the universe
at which the matter component dominates the cosmic fluid
(here we use $a_{i} \simeq 10^{-1}$ or $z_{i}\simeq 10$). 
Then the linear
growth factor normalized to unity at the present epoch
is $D(a)=\frac{\delta_{m}(a,\gamma)}{\delta_{m}(1,\gamma)}$.

Lastly, let us conclude with a brief discussion regarding the 
functional form of $\Delta_{de}$. 
Generally, in order to investigate the evolution of the DE perturbations 
we need to solve the system of Eqs. (\ref{odedelta}) and (\ref{dodedelta}). 
However, one may easily check that the latter system contains a particular
solution, namely $\delta_{de}=(1+w)\delta_{m}$ 
(see also Bean \& Dore 2004; Abramo et al. 2007; Ballesteros \& Riotto 2008; 
Abramo et al. 2009). Also, analytical 
solutions under of specific conditions can be found  
in Sapone \& Kunz (2009) and Sapone \& Majerotto (2012).  
Note that the accelerated expansion of the universe 
poses the restriction $w<-1/3\Omega_{de0}$ which 
implies $\Delta_{de}<(3\Omega_{de0}-1)/3\Omega_{de0}$.
Based on the above arguments 
we find the following two interesting cases:
a) for the quintessence CLDE ($-1<w<-1/3\Omega_{de0}$) model with $\delta_{m}>0$ 
we can have overdense DE regions ($\delta_{de} > 0$) 
and b) we may have underdense DE regions ($\delta_{de}<0$: ''DE voids'') 
in the phantom DE regime as long as $\delta_{m}>0$.
In a forthcoming 
paper we attempt to investigate the general solution of the system 
(\ref{odedelta}) and (\ref{dodedelta}) and thus to provide a complete
classification of the DE structures. 
To this end, if we consider 
that the DE fluctuations exist in nature ($\delta_{de} \ne 0$) 
then they could potentially play a
role in the DE era defined as $z\le z_{\star}$, where
$z_{\star}$ is the redshift of matter-DE equality which is given by 
$z_{\star}=\left( \frac{\Omega_{m0}}{\Omega_{de0}}\right)^{1/3w}-1$.

\begin{figure}
\mbox{\epsfxsize=8.2cm \epsffile{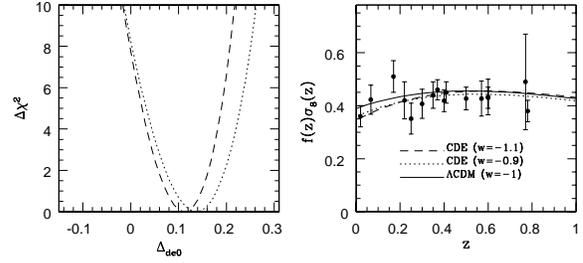}} \caption{{\em Left Panel:} 
The variance $\Delta \chi^{2}=\chi^{2}-\chi^{2}_{min}$
around the best fit $\Delta_{de0}$ value for the 
Quintessence ($w=-0.9$: dashed line) and Phantom 
($w=-1.1$: dotted line) models. 
{\em Right Panel:} Comparison of the observed and
theoretical evolution of the growth
rate $f(z)\sigma_{8}(z)$. Notice that for the curves 
we utilize the {\it Planck} priors provided by 
Spergel et al. (2013),
$(\Omega_{m0},\sigma_{8,\Lambda})=(0.30,0.818)$. 
The solid curve corresponds to the
concordance $\Lambda$CDM model.}
\end{figure}

\begin{figure}
\mbox{\epsfxsize=8.2cm \epsffile{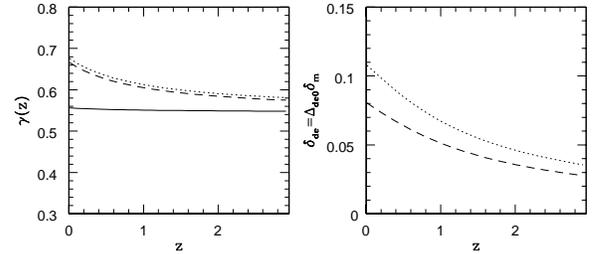}} \caption{The evolution 
of the growth index for the CLDE models (left panel).
In the right panel we provide the evolution of the 
dark energy perturbations.}
\end{figure}

\section{Observational constraints}
In the following we briefly present some details of the
statistical method and on the observational sample 
that we adopt in order to constrain the 
free parameter $\Delta_{de}$. 
We use the recent growth rate data
for which their combination parameter of the growth rate of structure,
$f(z)$, and the redshift-dependent 
rms fluctuations of the linear
density field, $\sigma_8(z)$,
is available as a function of redshift, $f(z)\sigma_{8}(z)$.
The total sample contains $N=16$ entries
(as collected by Basilakos et al. 2013 - see their Table 1 
and references therein) .

Notice that the $f\sigma_{8}$ estimator is almost a model-independent
way of expressing the observed growth history
of the universe (Song \& Percival 2009).
We use the standard $\chi^2$-minimization procedure, which in
our case it is defined as follows: 
\be
\label{Likel}
\chi^{2}({\bf p})=
\sum_{i=1}^{N} \left[ \frac{f\sigma_{8,\rm
      obs}(z_{i})-
f\sigma_{8}(z_{i},{\bf p})}
{\sigma_{i}}\right]^{2}
\ee
where $\sigma_{i}$ is the observed 1$\sigma$ uncertainty
and the theoretical growth-rate is given by:
$f\sigma_{8}(z,{\bf p})=\sigma_{8}D(z)\Omega_{m}(z)^{\gamma(z)}$.
The statistical vector ${\bf p}$ provides the
free parameters that enter in deriving the theoretical expectations.
For the case of constant $w$ it is
defined as: ${\bf p}=(\Omega_{m0},w,\Delta_{de0},\sigma_{8})$.
Since we are interested to check the DE perturbations at the present time
we restrict the likelihood analysis to the choice $\Omega_{m0}=0.30$ 
and $\sigma_{8,\Lambda}=0.818\left(0.30/\Omega_{m0}\right)^{0.26}$
provided by the {\em Planck} analysis of Spergel et al. (2013). 
Concerning the EoS we
set it either to $w=-1.1$ (phantom Ade et al. 2014 or 
to $w=-0.9$ (quintessence). Also, in order to use the $\sigma_{8}$
prior properly along the DE models we rescale the values of   
$\sigma_{8}$ by 
$\sigma_{8}=\frac{\delta_{m}(z=0)}{\delta_{m,\Lambda}(z=0)} \sigma_{8,\Lambda}$.
In that case the statistical vector includes only one free
parameter, ${\bf p}=\Delta_{de0}$.

Below we briefly discuss the main statistical results:
\noindent
(a) {\it Homogeneous DE} ($\Delta_{de0}=0$):
For the concordance $\Lambda$ cosmology we find that the 
theoretical $(\gamma_{0\Lambda},\gamma_{1\Lambda})\simeq (0.556,-0.011)$ values 
reproduce the growth data with $\chi_{min}^{2}/dof \simeq 18.1/15$.
The number of degrees of freedom is $dof=N-k-1$, where $k$ is 
the number of the fitted parameters (in this case $k=0$).   
Evidently, the above value of the reduced $\chi_{min}^{2}$
suggests that the $\Lambda$CDM model 
can not simultaneously accommodate the 
{\em Planck} priors and the growth data (see also 
Macaulay, Wehus \& Eriksen 2013; Mehrabi et al. 2015).
For the quintessence and phantom DE models we find that
$(\gamma_{0},\gamma_{1})=(0.56,-0.012)$
with $\chi_{min}^{2} \simeq 18.4/15$
and $(\gamma_{0},\gamma_{1})=(0.554,-0.011)$ 
with $\chi_{min}^{2} \simeq 20.2/15$ respectively.
\noindent

(b) {\it Clustered DE} ($\Delta_{de0}\ne 0$, $k=1$):
In the left panel of Fig. 1 we present the variation of 
$\Delta \chi^{2}=\chi^{2}(\Delta_{de0})-\chi^{2}_{min}(\Delta_{de0})$
around the best $\Delta_{de0}$ fit value.
For the quintessence CLDE model (dashed line)
we find that the likelihood function of the growth data
peaks at $\Delta_{de0}=0.14\pm 0.04$
with $\chi^{2}_{min}/dof \simeq 8.2/14$ and thus we obtain
$(\gamma_{0},\gamma_{1})\simeq (0.677,-0.129)$ which are in 
tension with those of homogeneous quintessence DE (see above).
Alternatively, if we impose the particular solution 
$\Delta_{de0}=1+w$ and minimizing with respect to $w$
we find $w=-0.85\pm 0.05$ with $\chi^{2}_{min}/dof \simeq 7.7/14$.
The fitted value of $\Delta_{de0}=0.14$ is in agreement, within 
$1\sigma$ errors, with that of $\Delta_{de0}=1+w$, which implies 
that for the quintessence CLDE model the corresponding 
particular solution of the system (\ref{odedelta}) and (\ref{dodedelta})
$\delta_{de}=(1+w)\delta_{m}$ 
is consistent with the growth data. 
In the case of phantom CLDE (with $w=-1.1$) the best fit 
value is $\Delta_{de0}=0.11\pm 0.03$ (dotted line: left panel of Fig.1) 
with $\chi^{2}_{min}/dof \simeq 10.5/14$ and thus we obtain 
$(\gamma_{0},\gamma_{1})\simeq (0.667,-0.124)$. 

It is interesting to mention that 
for the quintessence model the above 
$\Delta_{de}$ measurements are in agreement with those 
predicted by previous studies. Indeed,  
Bean and Dore (2004) found that in the case of 
$w=-0.8$\footnote{In this study 
the effective sound speed lies in the interval 0-1.} the ratio 
$\Delta_{de}=\delta_{de}/\delta_{m}$ at the present 
time can reach up to $\sim 0.09$ (see their Fig.1). 

In the right panel of Fig.1, we plot the observed $f(z)\sigma_{8}(z)$
with the estimated growth rate
function [see $\Lambda$CDM - solid line, quintessence - dashed line and 
phantom - dotted line]. In the right panel of Fig.2 we present 
the evolution of the DE perturbations 
for the quintessence and phantom models, respectively.
Based on the aforementioned $\Delta_{de0}$ 
observational constraints and Eq.(\ref{Dz221}),
we can estimate the DE fluctuations at the present time\footnote{Inserting 
the values of $(\Omega_{m0},\gamma_{0},\gamma_{1})$ 
into Eq.(\ref{Dz221}) we find that the linear growth factor is 
$\delta_{m0}\simeq 0.774$ (for $w=-0.9$) and 
$\delta_{m0}\simeq 0.735$ (for $w=-1.1$) respectively.}. In particular, 
we find $\delta_{de0}=0.108\pm 0.031$ and $\delta_{de0}=0.081\pm 0.022$
for the quintessence the phantom models respectively.
In the left panel of Fig.2 we can also see our 
determination of $\gamma(z)$ as a function 
of redshift.
The comparison indicates that the growth index of the CLDE models 
strongly deviates with respect to that of the concordance $\Lambda$ cosmology.
As an example close to the present epoch 
the departure can be of the order of 
$(\frac{\gamma-\gamma_{\Lambda}}{\gamma_{\Lambda}})\% \sim 25\%$. 
Notice that Ballesteros and Riotto (2008) found a $\sim 5 \%$ 
while Sefusatti \& Vernizzi (2011) computed 
a $\sim 10\%$ difference (see footnote 5 in their paper). 
On the other hand using equations (1) and (2) 
Mehrabi et al. (2015) found a relative large amount of 
deviations, namely $\sim 20-30\%$.
These differences imply 
that the deviations of the growth index depend on the 
initial assumptions and limitations imposed in the general system of 
equations that govern the matter and DE fluctuations. This is something
which needs further investigation.

To complete our study we repeat the analysis by using those
priors derived by the {\it Planck} team (Ade et al. 2014), namely
$\Omega_{m0}=0.315$ and $\sigma_{8,\Lambda}=0.87(0.27/\Omega_{m0})^{0.3}$.
In brief we find: 
$\Delta_{de0}=0.19\pm 0.04$ with $\chi^2_{min}/dof\simeq 8.8/14$ 
and $\Delta_{de0}=0.16\pm 0.03$ with $\chi^2_{min}/dof\simeq 11.6/14$ 
for the quintessence and phantom CLDE models respectively.
Recently, the cosmological results of 
{\it Planck 2015} appeared in the literature (Ade et al. 2015) and 
thus we utilize the pair 
$(\sigma_{8, \Lambda},\Omega_{m0})_{Planck15}=(0.308,0.815)$, 
which is close to that provided by Spergel et al. (2013).
In this case, the likelihood function 
peaks at $\Delta_{de0}=0.15\pm 0.04$ with $\chi^2_{min}/dof\simeq 8.1/14$ 
and $\Delta_{de0}=0.12\pm 0.03$ with $\chi^2_{min}/dof\simeq 10.3/14$ 
for the quintessence and phantom CLDE models respectively.
We verify that if we increase the value $\Omega_{m0}$ and that of the rms 
fluctuation 
of the linear density field on 8$h^{-1}$Mpc scales 
then the corresponding likelihood function peaks 
at higher $\Delta_{de0}$ values.


Finally, in order to compare the above DE models we use the
{\em corrected} Akaike information criterion for small sample size 
(${\rm AIC}$; Akaike 1974; Sugiura 1978) which is defined
for the case of Gaussian errors, as:
$${\rm AIC}=\chi^2_{min}+2k+\frac{2k(k-1)}{N-k-1}.$$
A smaller value of AIC indicates a better model-data fit.
We observe that the values 
of AIC$_{\rm CLDE}$($\simeq 10.1-12.4$) are smaller than the
corresponding homogeneous DE values, namely AIC$_{\rm HDE}$$\simeq 18.1-20.2$ 
which indicate that the CLDE scenario appears to fit
better than the usual homogeneous DE the growth data 
(see also Mehrabi et al. 2015).
At this point it is interesting to 
mention that Nesseris \& Sapone (20015) proposed 
a model independent test in order 
to check possible departures from the $\Lambda$CDM cosmological 
model at perturbation level. 
Assuming that the growth data are free from systematics they 
found that in order to reproduce the growth data we need to deal either 
with a clustered DE scenario or with a modified gravity. 

\section{Conclusions}
To summarize, we derive a new formulation of the 
growth index of matter perturbations 
in the regime (sub-horizon) where the dark energy 
is allowed to cluster.
In this framework, we find that the
growth index is indeed affected by the DE perturbations.
In order to check 
the range of validity of such a scenario we perform a likelihood analysis 
using the recent growth data. 
We show that the CLDE models fit much better 
the observational data than those of homogeneous DE models.
With the next generation of surveys, based mainly on {\em Euclid} (see also 
Sapone et al. 2013) 
we will be able to check whether the DE perturbations do really 
exist in nature.

\section*{Appendix}
In this appendix we examine the linear equation that describes
the evolution of the total (matter and DE) perturbations. 
Specifically, using $w=const.$, the combination of 
Eqs. (\ref{odedelta}) and (\ref{dodedelta}) provides
\be
\label{odedeltatot}
\ddot{\delta}+ 2H\dot{\delta}=
\frac{3H^{2}}{2}(2+w)\left[\Omega_{m}\delta_{m}+\Omega_{de}\delta_{de}(1+3w)\right],
\ee
where $\delta\equiv \delta_{m}+\delta_{de}$ is the sum of matter and DE 
fluctuations. Obviously, for $(w,\delta_{de})=(-1,0)$ the above 
equation boils down to that of the concordance $\Lambda$ cosmology.
Now, considering that the overall growth rate of clustering 
is given by $f=d{\rm ln}\delta/d{\rm ln} a\simeq \Omega_{m}^{\gamma(z)}(z)$ and  
following simultaneously the procedure described in section 2 we arrive at 
\be
\label{fzz333}
a\frac{df}{da}+f^{2}+\left( \frac{1}{2}-\frac{3}{2}w\Omega_{de}\right)f
= 
\frac{3(2+w)\left[\Omega_{m}+ (1+3w)\Delta_{de} \Omega_{de} \right]}
{2(1+\Delta_{de})}
\ee
or
\be
\label{Polll}
-(1+z)\gamma^{\prime}{\rm ln}(\Omega_{m})+\Omega_{m}^{\gamma}+
3w\Omega_{de}(\gamma-\frac{1}{2})+\frac{1}{2}=
\frac{3}{2}\Omega_{m}^{1-\gamma}{\tilde X} 
\ee
where prime denotes derivative with respect to redshift and 
\be \label{xxxx}
{\tilde X}(z)=\frac{(2+w)X(z)}{1+\Delta_{de}(z)}.
\ee 
We remind the reader that the function $X(z)$ is given by Eq.(\ref{xxx}) and 
$\gamma(z)=\gamma_{0}+\gamma_{1}y(z)$ with $y(z)=z/(1+z)$.
Therefore, replacing $X_{0}=X(z=0)$ with ${\tilde X}_{0}={\tilde X}(z=0)$ 
in Eq.(\ref{Poll2}) we have $\gamma_{1}$ in terms of $\gamma_{0}$.
Finally, repeating our statistical analysis 
we find the following results: 

In the case of Spergel et al. (2013) {\it Planck} priors 
$(\Omega_{m0},\sigma_{8,\Lambda})=(0.30,0.818)$:
\begin{itemize}
\item for the quintessence model: 
$\Delta_{de0}=0.13\pm 0.04$
with $\chi^{2}_{min}/dof \simeq 8.2/14$, 
AIC$\simeq 10.2$ and 
$(\gamma_{0},\gamma_{1})\simeq (0.671,-0.123)$.

  \item for the phantom model: 
$\Delta_{de0}=0.10\pm 0.04$
with $\chi^{2}_{min}/dof \simeq 10.5/14$, 
AIC$\simeq 12.5$ and 
$(\gamma_{0},\gamma_{1})\simeq (0.674,-0.131)$.

\end{itemize}

In the case of the {\it Planck 2015} priors (Ade et al. 2015)
$(\Omega_{m0},\sigma_{8,\Lambda})=(0.308,0.815)$:
\begin{itemize}
\item for the quintessence model: 
$\Delta_{de0}=0.15\pm 0.04$
with $\chi^{2}_{min}/dof \simeq 8.1/14$, 
AIC$\simeq 10.1$ and 
$(\gamma_{0},\gamma_{1})\simeq (0.689,-0.141)$.

\item for the phantom model: 
$\Delta_{de0}=0.11\pm 0.04$
with $\chi^{2}_{min}/dof \simeq 10.3/14$, 
AIC$\simeq 12.3$ and 
$(\gamma_{0},\gamma_{1})\simeq (0.684,-0.141)$.
\end{itemize}
Obviously the above results are in agreement within $1\sigma$ 
with those presented in  section 3.

\section*{Acknowledgments}
SB acknowledges support by the Research Center for Astronomy of
the Academy of Athens in the context of the program  ``{\it Tracing
the Cosmic Acceleration}''. SB are also grateful to the
Department ECM (Universitat de Barcelona) for the hospitality and
support when this work was being finished.

\end{document}